\documentclass[twocolumn,showpacs,aps,prl,superscriptaddress]{revtex4}

\usepackage{graphicx}
\usepackage{dcolumn}
\usepackage{amsmath}
\usepackage{epsfig}

\usepackage{lscape}
\usepackage{rotating}
\usepackage{wrapfig}
\usepackage{psfrag}
\usepackage{color}
\usepackage{rotate}
\usepackage{amssymb}

\input babarsym
 \def\eg         {\ensuremath{E_{\gamma} }\xspace}

 \def\mup        {\ensuremath{\mu_{\pi}}\xspace}
 \def\mupisq    {\ensuremath{\mu_{\pi}^2}\xspace}

 \def\mb         {\ensuremath{m_{b} }\xspace}

 \def\acp        {\ensuremath{A_{CP}}\xspace}

 \def\bsg        {\ensuremath{b \to s \gamma}\xspace}
 \def\bdg        {\ensuremath{b \to d \gamma}\xspace}

 \def\bxsg       {\ensuremath{B \to X_{s} \gamma}\xspace}
 \def\bxdg       {\ensuremath{B \to X_{d} \gamma}\xspace}
 
 \def\bxg        {\ensuremath{B \to X \gamma}\xspace}

\def\aveDelta#1 {\ensuremath{\langle \Delta_{total}#1 \rangle}\xspace}

\def\ecut {\ensuremath{E_{\rm cut}}}
 
 \def\meg {\ensuremath{\langle E_{\gamma} \rangle}}
 
 \def\vegs {\ensuremath{\langle (E_{\gamma} -\meg)^2 \rangle}}

\def\B      {\ensuremath{B}\hbox{ }}
\def\Bp      {\ensuremath{B^{+}}\hbox{ }}

\newcommand{\beq}{\begin{equation}}
\newcommand{\beqa}{\begin{eqnarray}}
\newcommand{\beqn}{\begin{eqnarray}}
\newcommand{\eeq}{\end{equation}}
\newcommand{\eeqa}{\end{eqnarray}}
\newcommand{\eeqn}{\end{eqnarray}}
\makeatletter
\def\slash#1{{\mathpalette\c@ncel{#1}}} % TeXbook, bottom of p360
\makeatother
\newcommand{\BABARPubYear}    {07}
\newcommand{\BABARPubNumber}  {067}

\newcommand{\SLACPubNumber} {13021}
\newcommand{\LANLNumber} {xxx}

\def\figurebox#1#2#3{%
    \def\arg{#3}%
    \ifx\arg\empty
    {\hfill\vbox{\hsize#2\hrule\hbox to #2{\vrule\hfill\vbox to 
     #1{\hsize#2\vfill}\vrule}\hrule}\hfill}%
    \else
    {\hfill\epsfbox{#3}\hfill}%
    \fi}

\begin{document}

\preprint{hep-ex/\LANLNumber}

\begin{flushleft}
\babar-PUB-\BABARPubYear/\BABARPubNumber\\ 
SLAC-PUB-\SLACPubNumber\\
arXiv:0711.4889 [hep-ex]
\end{flushleft}

\title{
{\large Measurement of the \bxsg Branching Fraction and \\ 
Photon Energy Spectrum using the Recoil Method} 
}

% Dummy author list; contact PubBoard Chair for current author list
%% author list as of 04-Sep-2007 (556 authors)
%
\author{B.~Aubert}
\author{M.~Bona}
\author{Y.~Karyotakis}
\author{J.~P.~Lees}
\author{V.~Poireau}
\author{X.~Prudent}
\author{V.~Tisserand}
\author{A.~Zghiche}
\affiliation{Laboratoire de Physique des Particules, IN2P3/CNRS et Universit\'e de Savoie, F-74941 Annecy-Le-Vieux, France }
\author{J.~Garra~Tico}
\author{E.~Grauges}
\affiliation{Universitat de Barcelona, Facultat de Fisica, Departament ECM, E-08028 Barcelona, Spain }
% ...
\collaboration{The \babar\ Collaboration}
\noaffiliation

% It is always \today, today, but you may specify any date with \date.
\date{\today}

\begin{abstract}

We present a measurement of the branching fraction and photon energy
spectrum for the decay \bxsg using data from the \babar\ experiment.
The data sample corresponds to an integrated luminosity of 210
fb$^{-1}$, from which approximately 680\,000 \BB events are tagged by
a fully reconstructed hadronic decay of one of the $B$ mesons. In the
decay of the second $B$ meson, an isolated high--energy photon is
identified. 
We measure $\mathcal{B} (\bxsg)= (3.66 \pm 0.85_{\rm stat} \pm 0.60_{\rm syst})
\times 10^{-4}$ for photon energies \eg above 1.9\gev in the $B$ rest frame.  
From the measured spectrum we calculate the first and second moments for
different minimum photon energies, which are used to extract the heavy-quark 
parameters $m_{\rm b}$ and $\mu_{\pi}^2$.  In addition,
measurements of the direct \CP asymmetry and isospin asymmetry are
presented.

\end{abstract}

%\pacs{13.25.Hw, 12.15.Hh, 11.30.Er}% PACS, the Physics and Astronomy Classification Scheme.
\pacs{13.20.He, 13.30.Ce, 12.39.Hg}
%11.30.Er 	Charge conjugation, parity, time reversal, and other discrete symmetries
%12.15.Hh 	Determination of Kobayashi-Maskawa matrix elements
%12.39.Hg Heavy quark effective theory
%13.20.-v Leptonic, semileptonic and radiative decays of mesons
%13.20.He decays of bottom mesons
%13.25.Hw same as above!!
%13.30.Ce 	Leptonic, semileptonic, and radiative decay
\maketitle

\section{Introduction}
\label{sec:intro}

We present measurements of the branching fraction and photon energy
spectrum of the rare radiative penguin decay \bxsg using $\FourS \ra
\BB$ events. We use a new technique where one of the $B$ mesons
(called the tag $B)$ decays to hadrons and is fully reconstructed.
This approach allows for the determination of the charge, flavor and 
momentum of both of the $B$ mesons, and thus the photon spectrum can
be determined in the rest frame of the signal $B$.  The method 
results in an improved purity for the signal sample,
allows separate 
measurements for charged and neutral $B$ mesons and enables the 
measurement of the direct \CP asymmetry \acp.
This approach is complementary to those used in previous studies
\cite{Chen:2001fj,Koppenburg:2004fz,BABARSEMI,BABARINCL}
and incurs different systematic uncertainties.

In the Standard Model (SM), the  decay \bsg
proceeds via a flavor-changing neutral current.  The decay is
sensitive to new physics through non-SM heavy particles entering at
the loop level \cite{Hurth:2003vb}.  Recent
next-to-next-to-leading-order calculations predict SM branching
fractions in the range $\mathcal{B}(\bxsg) = (3.0 - 3.5) \times 10^{-4}$ 
for $E_{\gamma}>1.6\gev$ with uncertainties that vary from 
$7\%$ to $14\%$~\cite{Misiak:2006zs,Becher:2006qw,Andersen:2006hr}.
Here $E_{\gamma}$ is the energy of the signal photon in the rest frame of
the $B$ meson, and the cutoff is chosen to avoid non-perturbative effects
at lower energies.  The current world average measured branching
fraction is $\mathcal{B}(\bxsg) = (3.55 \pm 0.26) \times 10^{-4}\,
(E_{\gamma}>1.6\gev)$~\cite{hfag2007,Buchmuller:2005zv}.
The moments of the photon energy spectrum are sensitive to the
Heavy Quark Expansion parameters \mb and \mupisq
%~\cite{Kagan:1998ym,Benson:2004sg,Lange:2005yw}
, related to the mass
and momentum of the \b quark within the $B$ meson~\cite{Kapustin:1995nr}.  
Improved measurements of these parameters can be used to reduce the 
uncertainty in the CKM matrix elements $\Vcb$ and
$\Vub$~\cite{hfag2007,Buchmuller:2005zv}.

The measurements presented here are based on a sample of 232 million \BB\ 
pairs collected  on the \FourS\ resonance by the \babar\ 
detector~\cite{Aubert:2001tu} at the PEP-II  asymmetric-energy  $e^+e^-$ 
storage ring operating at SLAC, corresponding to an integrated luminosity 
of $210\,\mbox{fb}^{-1}$.
After reconstruction of the tag $B,$ the remaining particles in the event
are assigned to the second $B$ (the signal $B)$ and events
containing a high-energy photon
are selected.  The signal process $\bxg$ at this stage is taken to
mean events from either $\bsg$ or $\bdg$ decays; the small
contribution from $\bdg$ is subtracted at the end of the analysis.
The sample also includes background from continuum (non-\BB) events
and \BB events in which the tag $B$ is misreconstructed. These are
subtracted by means of a fit to the beam-energy-subsituted mass
(defined below) of the tag $B$.

The remaining background events, where the photon candidate is not from
the signal process (e.g., a photon from a $\pi^0$ or $\eta$ decay), are
subtracted using a Monte Carlo (MC) model based on
\evtgen~\cite{Lange:2001uf} and \geantf~\cite{Agostinelli:2002hh}.
The MC predictions are
scaled to data in the low $E_{\gamma}$ region, where the signal
contribution is very small.  This allows a reliable measurement for
photon energies $E_{\gamma} > 1.9\gev$. Finally, to compare with other
experiments and predictions, the measured rate is extrapolated using
theoretical models to give the rate for $E_{\gamma} > 1.6\gev$.

This measurement is currently limited by
statistics, and furthermore, the dominant systematic errors are of the
type that should decrease with a larger data sample. 
Therefore, the approach followed here is expected to provide an
increasingly competitive level of precision when applied to the
larger data sample currently being collected by the \babar\
experiment.

%%%%%%%%%%%%%%%%%%%%%%%%%%%%%%%%%%%%%%%%%%%%%%%%%%%%%%%%%%%%%%%%%%
\section{Event selection}
\label{sec:selection}
%%%%%%%%%%%%%%%%%%%%%%%%%%%%%%%%%%%%%%%%%%%%%%%%%%%%%%%%%%%%%%%%%%

Using 1114 exclusive hadronic decay channels \cite{Aubert:2003zw}, 
which represent about 5\% of the total decay width of the $B^0$ and $B^+$ 
mesons, we identify events in which one of the two B mesons is fully 
reconstructed.
The kinematic consistency of the tag $B$ candidates is checked with
two variables, the beam-energy-substituted mass
$\mes = \sqrt{s/4 - \vec{p}_B^2}$, and the energy difference $\Delta E = E_B -
\sqrt{s}/2$, where $s$ is the total energy squared in the
center-of-mass (c.m.) frame, and $E_B$ and $\vec{p}_B$ are the the
c.m.\ energy and momentum of the tag $B$ candidate. We require
$|\Delta E| \leq 60\mev$, a window of approximately $\pm 3 \sigma$. 

Those particles in the event that are not reconstructed as part of the
tag $B$ are regarded as coming from the signal $B$.  Among these
particles we require an isolated photon candidate with energy
$E_{\gamma} > 1.3\gev$ in the $B$ frame. To ensure a well
reconstructed photon, we require the electromagnetic shower to lie well
within the calorimeter acceptance and to satisfy isolation and shower
shape requirements.

The background events consist of non-signal $B$ decays and continuum
background from \uubar, \ddbar, \ssbar and \ccbar events.  The
continuum events are suppressed by using a Fisher
discriminant that combines 12 variables related to
the different event decay topologies of \BB and continuum events.
These include event-shape variables such as the thrust, as well as
information on the energy flow relative to the direction of the
candidate signal photon.

To discriminate against photons from \piz and $\eta$ decays,
we combine the signal candidate photon with any other photon in the
event associated with the signal $B$.  The event is vetoed if the
pair's invariant mass is consistent with a \piz or $\eta$.
Furthermore, the event is rejected if the candidate photon combined with
a $\pi^{\pm}$
is consistent with a $\rho^{\pm} \ra \pi^{\pm}\piz$ decay
assuming that the second photon from the \piz decay is lost.

%%%%%%%%%%%%%%%%%%%%%%%%%%%%%%%%%%%%%%%%%%%%%%%%%%%%%%%%%%%%%%%%%%
\section{Fit of signal rates}
\label{sec:fit}
%%%%%%%%%%%%%%%%%%%%%%%%%%%%%%%%%%%%%%%%%%%%%%%%%%%%%%%%%%%%%%%%%%

The distribution of \mes for the selected events has a peak around the
mass of the $B$ meson, corresponding to correctly reconstructed
\BB events, and a broad background component that stems from non-\BB 
and misreconstructed \BB events.  The peak is modeled with a
Crystal Ball (CB) function~\cite{Skwarnicki:1986xj}.  This contains
two parameters that correspond to the mean and width of the Gaussian
core and two additional parameters that describe a power-law tail
extended to masses below the core region.  The non-peak background
term is described with an ARGUS function~\cite{Albrecht:1987nr}.

Applying the selection criteria outlined above yields approximately
7\,700 events. 
We divide the event sample into 14 intervals of
photon energy, each 100\mev wide, spanning the range 1.3 to 2.7\gev. 
In  each interval, we extract the number of peak events with a binned 
maximum likelihood fit to the \mes distribution.

\begin{figure*}[htb]
  \begin{center}
  \includegraphics{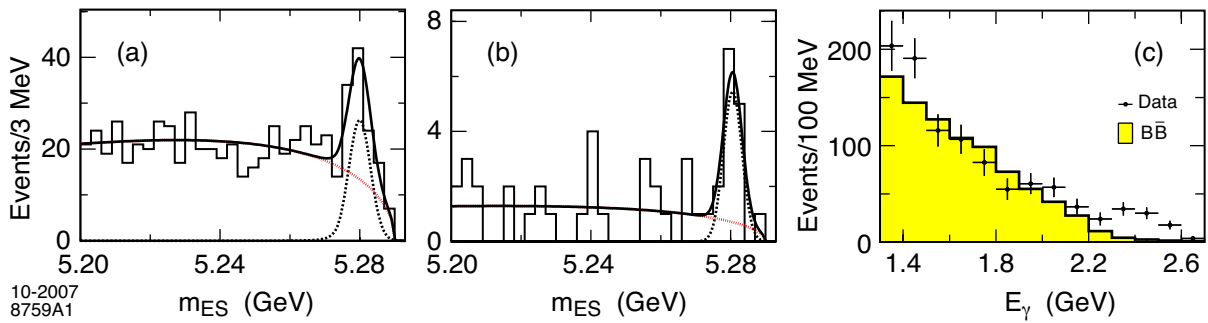}
  \caption{\label{fig:mesfits} Fits to the distribution of the 
beam-energy-substituted mass $\mes$ for two \eg regions.  
The dashed curve shows the CB term and the dotted 
curve is the ARGUS term, corresponding to $B$ and non-$B$
events, respectively;  the solid curve is their sum.
(a)  $1.6\gev < \eg < 1.7\gev$ for the charged $B$ sample.
(b)  $2.3\gev < \eg < 2.4\gev$ for the neutral $B$ sample.
(c) The measured numbers of $B$ events as a function of photon
energy.  The points are from data; the histogram is from a \BB MC
sample which excludes the signal decay \bxg.
}
  \end{center}
\end{figure*}

The limited size of the data sample means that it is not possible to fit 
all of the parameters related to the shape of the CB and ARGUS
functions individually in separate intervals of photon energy.  One
expects, however, a smooth variation of the shapes as a function of
$E_{\gamma}$.  To impose this smoothness, a simultaneous fit of the
\mes distributions for all of the photon-energy intervals is carried
out.  The variation of the shape parameters with photon energy is described
by polynomials, whose orders are the lowest possible that allow an
adequate modeling of the data.
Examples of the \mes distributions and results of the simultaneous fit
are shown in Fig.~\ref{fig:mesfits}.  The global $\chi^2$ is 330 for
the charged $B$ sample and 357 for the neutral sample, both for 387
degrees of freedom.

The measured numbers of $B$ events are shown in Fig.~\ref{fig:mesfits}~(c) 
as a function of photon energy.  The points are from data; the
solid histogram is from a \BB MC sample that
excludes the signal decay \bxg.  
%CHF
Due to the large background at low energy the signal region is defined as 
$E_{\gamma} > 1.9\gev$. This choice was optimized in MC studies.
The MC prediction has been
scaled by fitting to the data region between $1.3 \, < E_{\gamma}
< 1.9\gev$, taking into account the small contribution from \bxg
decays in that region. For $\eg > 1.9\gev$, we observe $119 \pm 22$ 
\bxg signal events over a \BB background of $145 \pm 9$ events.

For $1.3 < \eg < 1.9\gev$ a comparison of the data and background
gives a $\chi^2$ of 9.7 for five degrees of freedom.  The probability
to observe a value at least this great is 8.4\%.
Our estimate of the systematic uncertainty in the background (described
below) is in fact smaller than the observed data-background difference;
therefore we regard this difference primarily as a statistical
fluctuation.

To determine the partial branching fractions, we require
the total number of \BB events in the sample after selection
of the tag $B$ candidates. 
In a procedure analogous to that described for the \mes fits in bins of
$E_{\gamma}$, we divide the data into four intervals of estimated tag $B$
candidate purity and perform a simultaneous fit of the \mes distributions.
We obtain approximately 680\,000 \BB events corresponding to an efficiency 
of 0.3\%.

%%%%%%%%%%%%%%%%%%%%%%%%%%%%%%%%%%%%%%%%%%%%%%%%%%%%%%%%%%%%%%%%%%%%
\section{Determining the photon spectrum}
\label{eq:corrections}
%%%%%%%%%%%%%%%%%%%%%%%%%%%%%%%%%%%%%%%%%%%%%%%%%%%%%%%%%%%%%%%%%%%%

The differential decay rate $(1/\Gamma_B) (d\Gamma/dE_{\gamma})$ is
measured in bins of the ($B$-frame) photon energy for $E_{\gamma} > 1.9\gev$ 
up to the kinematic limit at 2.6\gev.
It is estimated for the $i$th bin as

\begin{equation}
\label{eq:estimator} 
\frac{1}{\Gamma_B} 
\frac{d\Gamma_i} {dE_{\gamma}}
= \frac{N_i - b_i} {\varepsilon_i N_B } \;,
\end{equation}

\noindent where $N_i$ is the number of $B$ events in the bin, $b_i$ is the 
number of $B$ mesons from decays other than \bxg, $N_B$ is the total
number of $B$ mesons in the sample, and $\varepsilon_i$ is the
efficiency, which corrects for both acceptance and bin-to-bin
resolution effects. The values $b_i$ are determined by means
of a simultaneous fit to the $\mes$ distributions as described
previously, using a sample of MC data consisting of \BB
events excluding the signal decay \bxg.
As the differential decay rate is normalized
using the total width of the $B$ meson, $\Gamma_B$, the integral of
(\ref{eq:estimator}) over all photon energies yields the branching
fraction.
To evaluate the selection efficiency $\varepsilon_i$, we model the
signal photon energy spectrum using the kinetic scheme
\cite{Benson:2004sg} with $m_b = 4.60\gev$ and $\mupisq = 0.4 \gev^2$.
The value of $\varepsilon_i$ is determined from
 
\begin{equation}
\label{eq:eff}
\varepsilon_i = \frac{N_{{\rm found},i}/ N_{\rm sim} }
{N_{{\rm true},i}/N_{\rm gen}}  \; C_{\rm tag} \;,
\end{equation}
 
\noindent where $N_{{\rm found},i}$ is the number of events found
in a MC sample of \bxsg with
detector simulation and $N_{\rm sim}$ is the number of events in the
simulated sample.  These quantities are found using the same fit
procedure as applied to the real data for $N_i$ and $N_B$. In the
denominator of (\ref{eq:eff}), $N_{{\rm true},i}$ is the true number
of events with photon energies in bin $i$ and $N_{\rm gen}$ is the
total number of events generated.  These values are determined
using the event generator for \bxsg decays only,
without detector simulation. The factor $C_{\rm tag}$,
estimated using the MC model, corrects for the small dependence of
the probability to find a tag $B$ on the presence of a \bxg final state.
The efficiency increases roughly linearly with photon energy, and is
approximately 30\% (65\%) for $E_{\gamma} = 1.9\gev$ $(2.6\gev)$.

To compare with other results we subtract the \bxdg component from the
differential decay rates using the Standard Model prediction (for the
\CP and isospin asymmetries discussed below, however, we do not make this
correction). The values $\mathcal{B}(B \rightarrow X_d \gamma)$ and
$\mathcal{B}(B \rightarrow X_s\gamma)$ are in the ratio
$|V_{td}/V_{ts}|^2$ assuming the same efficiency for the two
categories of events.  Therefore, the branching ratio is lowered by
$(4.0 \pm 0.4) \%$ \cite{Hurth:2003dk,Charles:2004jd}.

%%%%%%%%%%%%%%%%%%%%%%%%%%%%%%%%%%%%%%%%%%%%%%%%%%%%%%%%%%%%%%%%%%%%%%
\section{Systematic Uncertainties}
\label{sec:systematics}
%%%%%%%%%%%%%%%%%%%%%%%%%%%%%%%%%%%%%%%%%%%%%%%%%%%%%%%%%%%%%%%%%%%%%%

There are four main sources of systematic uncertainty, which are
summarized in Table~\ref{tab:results}: modeling of the \BB background,
the $\mes$ fits, detector response and dependence on the \bxsg signal
model. In addition there is an uncertainty from the subtraction of the
$\bxdg$ contribution.

After subtraction of the non-peak background using the \mes
distribution, the remaining background is mainly composed of \BB events
with the selected photon coming from a \piz or $\eta$ decay.  
Photons from \piz account for $55\%$ to $65\%$
depending on \eg and the charge of the tag $B,$ while the contribution from
$\eta$ mesons varies from $18\%$ to $29\%$.  The remaining backgrounds
include fake photons from $\bar{n}$ annihilation, real photons from
bremsstrahlung or from $\omega$ decays, and electromagnetic 
showers from $e^{\pm}$ misidentified as photons.
As the MC prediction for the \BB background is scaled to the data at
low energy, there is no uncertainty stemming from the absolute rate,
but rather only from the shape of the distribution as a function of
$E_{\gamma}$.
The uncertainty from the inclusive \piz and $\eta$ spectra is
investigated by using \eg dependent correction factors for the \piz and
$\eta$ yields from a large control sample of \bxg
candidate events, obtained using a lepton tag. These
correction factors are typically around 5\% for \piz yields while they
can be up to 30\% for $\eta$ yields.  
The remaining backgrounds have a roughly linear slope with $E_{\gamma}$;
this is varied by $\pm 30\%$.
We use the difference obtained with the modified MC compared to the
standard MC simulation as a systematic uncertainty.

To assess the uncertainty related to the parameterization
chosen for the $\mes$ fit, additional coefficients are
introduced that allow linear or higher-order dependence of the CB and 
ARGUS function shape parameters on the photon energy.  The
maximum variation in the fitted rates is taken as the
systematic uncertainty.  A similar set of variations for the
dependence of the shape parameters on the $B$ meson purity is carried out
for the $\mes$ fits used to determine the total number of $B$
mesons in the data sample.
To allow for a small peaking component in the distribution of \mes
from \Bpm decays reconstructed as \Bz (\Bzb) decays and vice versa, we 
remove these events from the MC sample and take the difference 
in the result as a systematic uncertainty.

The uncertainties related to the detector modeling and event
reconstruction are estimated by comparing MC simulations of
track and photon efficiencies as well as particle identification
efficiencies with data control samples.  From these comparisons
we estimate corresponding systematic errors, which are in all cases
small compared to other uncertainties.

To assess the uncertainty in the efficiency due to the assumed shape of
the $E_{\gamma}$ spectrum, we vary $m_b$ and $\mupisq$ in the kinetic
scheme by $\pm0.1\gev$ and $\pm0.1\gev^2$, respectively.  
These variations are large compared to the uncertainties in the
world average~\cite{Buchmuller:2005zv} in order to cover alternative 
Ans\"atze for the 
the heavy quark distribution function~\cite{Kagan:1998ym,Lange:2005yw}.
They also account for uncertainties related to the small rate of \bxg 
decays expected below 1.9\gev.

%%%%%%%%%%%%%%%%%%%%%%%%%%%%%%%%%%%%%%%%%%%%%%%%%%%%%%%%%%%%%%%%%%%%%%%%%%%%%
\section{Results}
\label{sec:results}
%%%%%%%%%%%%%%%%%%%%%%%%%%%%%%%%%%%%%%%%%%%%%%%%%%%%%%%%%%%%%%%%%%%%%%%%%%%%%

\begin{figure}[t]
\begin{center}
\includegraphics{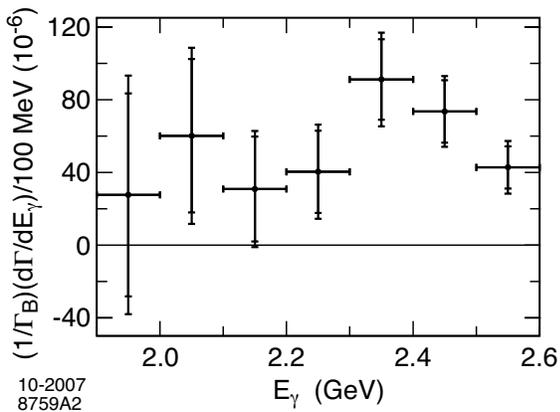}
\end{center}
\caption{
The partial branching fractions $(1/\Gamma_B) (d\Gamma/dE_{\gamma})$
with statistical (inner) and total (outer) uncertainties.
}
\label{fig:pbf}
\end{figure}

The partial branching fractions $(1/\Gamma_B) (d\Gamma/dE_{\gamma})$
are shown in Fig.~\ref{fig:pbf} after all corrections.
The inner error bars show the statistical uncertainties.  The outer
error bars show the quadratic sum of the statistical and systematic
terms.  By integrating the spectrum, we obtain ${\mathcal B}(\bxsg) =
(3.66 \pm 0.85_{\rm stat} \pm 0.60_{\rm syst}) \times 10^{-4}$.  The
results for the differential decay rate and for the moments of the
photon energy spectrum for various minimum photon energies $E_{\rm cut}$
are given in Table~\ref{tab:results}. The branching fraction for larger 
values of $E_{\rm cut}$ and the correlations between the measurements 
are given in Tables~\ref{tab:bfresults}-\ref{correlsys}.  
Our results are in good agreement with those presented in
Refs.~\cite{Chen:2001fj,Koppenburg:2004fz,BABARSEMI,BABARINCL}.

\begin{table}
\begin{center}
\caption[]{\label{tab:results} Results for the differential decay rate $(1/\Gamma_B) (d\Gamma/dE_{\gamma})$
and moments of the photon spectrum with statistical and 
systematic errors. The major contributions to the systematic 
uncertainties are also listed: (a) background modeling, (b) \mes fit parameterization, (c) detector response, (d) \bxsg model.
}
\begin{tabular}{l l l l l l l l}
\hline\hline
\multicolumn{8}{c}{ $(1/\Gamma_B) (d\Gamma/dE_{\gamma}) \, (10^{-4})$}\\
 $E_{\gamma}$\,(\gev) & Value  &  $\sigma_{\rm stat}$ & $\sigma_{\rm  syst}$ & (a) & (b)  & (c)    & (d) \\\hline
   1.9-2.0	&	0.28	&	0.56	&	0.34	&	0.26	&	0.13	&	0.19	&	0.03	\\
   2.0-2.1	&	0.60	&	0.42	&	0.24	&	0.18	&	0.12	&	0.08	&	0.05	\\
   2.1-2.2	&	0.31	&	0.29	&	0.14	&	0.11	&	0.06	&	0.03	&	0.03	\\
   2.2-2.3      &	0.40	&	0.23	&	0.13	&	0.07	&	0.05	&	0.09	&	0.03	\\
   2.3-2.4	&	0.91	&	0.22	&	0.13	&	0.07	&	0.08	&	0.05	&	0.06	\\
   2.4-2.5	&	0.74	&	0.17	&	0.09	&	0.05	&	0.05	&	0.02	&	0.05	\\
   2.5-2.6	&	0.43	&	0.12	&	0.09	&	0.03	&	0.03	&	0.07	&	0.04	\\\hline
\multicolumn{8}{c}{ \meg \, (\gev)}\\
 $E_{\gamma}$\,(\gev) & Value  &  $\sigma_{\rm stat}$ & $\sigma_{\rm  syst}$ & (a) & (b)  & (c)    & (d) \\\hline
1.9-2.6	&	2.289	&	0.058	&	0.027	&	0.018	&	0.019	&	0.009	&	0.002	\\
2.0-2.6	&	2.315	&	0.036	&	0.019	&	0.013	&	0.011	&	0.009	&	0.001	\\
2.1-2.6	&	2.371	&	0.025	&	0.009	&	0.007	&	0.005	&	0.003	&	0.001	\\
2.2-2.6	&	2.398	&	0.016	&	0.004	&	0.003	&	0.003	&	0.001	&	0.000	\\
2.3-2.6	&	2.427	&	0.010	&	0.006	&	0.000	&	0.001	&	0.005	&	0.000	\\\hline
\multicolumn{8}{c}{ \vegs \, ($\gev^2$)}\\
 $E_{\gamma}$\,(\gev) & Value  &  $\sigma_{\rm stat}$ & $\sigma_{\rm  syst}$ & (a) & (b)  & (c)    & (d) \\\hline
1.9-2.6	&	0.0334	&	0.0124	&	0.0062	&	0.0040	&	0.0025	&	0.0037	&	0.0013	\\
2.0-2.6	&	0.0265	&	0.0057	&	0.0024	&	0.0018	&	0.0010	&	0.0007	&	0.0011	\\
2.1-2.6	&	0.0142	&	0.0037	&	0.0013	&	0.0009	&	0.0005	&	0.0004	&	0.0006	\\
2.2-2.6	&	0.0092	&	0.0015	&	0.0010	&	0.0002	&	0.0002	&	0.0009	&	0.0003	\\
2.3-2.6	&	0.0059	&	0.0007	&	0.0003	&	0.0000	&	0.0000	&	0.0003	&	0.0002	\\\hline\hline
\end{tabular}
\end{center}
\end{table}

We also measure the isospin asymmetry $\Delta_{0-}$,

\begin{equation}
\label{eq:isospin}
\Delta_{0-} = \frac{\Gamma(\Bzb\rightarrow X_{s,d} \, \gamma) - 
\Gamma(\Bm \rightarrow X_{s,d} \, \gamma)}
{\Gamma(\Bzb\rightarrow X_{s,d} \, \gamma)+
\Gamma(\Bm\rightarrow X_{s,d} \, \gamma)} \;,
\end{equation}

\noindent where inclusion of charge conjugate modes is implied.
%This asymmetry is expected to be small for \bxsg, with isospin
%breaking only occurring at higher orders in the heavy quark
%expansion.  
%CHF
It has been argued that enhanced power corrections to the \bxsg rate 
could also lead to values of $\Delta_{0-}$ as large as $+10\%$~\cite{Lee:2006wn}.
Therefore, experimental measurements of $\Delta_{0-}$ can help determine the size of these 
effects and hence reduce the theoretical uncertainty on the total rate.
%CHF
To obtain decay rates from the branching fractions we use the 
$B$ meson lifetimes: $\tau(\Bz)= 1.530\pm
0.008$ ps and $\tau(\Bp)= 1.638\pm 0.011$ ps~\cite{pdg2006}.  For
photon energies greater than 2.2
\gev, we obtain $\Delta_{0-} = -0.06 \pm 0.15_{\rm stat} \pm
0.07_{\rm syst}$.

The direct \CP asymmetry \acp ,
\begin{equation}
\label{eq:acp}
\acp = \frac{\mathcal{B}(B\rightarrow X_{s,d} \, \gamma) - 
\mathcal{B}(\Bb \rightarrow X_{s,d} \, \gamma)}
{\mathcal{B}(B \rightarrow X_{s,d} \, \gamma) + 
\mathcal{B}(\Bb \rightarrow X_{s,d} \, \gamma)}\frac{1}{1-2\omega} \,,
\end{equation}
is measured by splitting the tag sample into $B$ and \Bb mesons.
The dilution factor $\frac{1}{1-2\omega}$ accounts for the mistag fraction $\omega$,
here simply the time integrated \Bz mixing probability
of $\chi_d = 0.188 \pm 0.003$~\cite{pdg2006} multiplied by the fraction of \Bz events in 
the total data sample.
$\acp$ can be significantly enhanced by new physics~\cite{Hurth:2003dk}
while in the SM it is predicted to be around $10^{-9}$~\cite{Soares:1991te,Hurth:2001yb}.
We obtain a value of $\acp = 0.10 \pm  0.18_{\rm stat} \pm  0.05_{\rm syst}$ 
for photon energies above 2.2\gev.

For both $\Delta_{0-}$ and \acp, a photon energy cutoff of 2.2\gev is
chosen because it facilitates comparison with previous results and
minimizes the total uncertainty.  Our results are in good agreement
with previous
measurements~\cite{Coan:2000pu,Aubert:2004hq,BABARSEMI,BABARINCL,Nishida:2003yw}.

Finally, we use heavy quark expansions in the kinetic
scheme~\cite{Benson:2004sg} and our measurements of the $E_{\gamma}$
moments to determine the parameters $m_{b}$ and $\mu_{\pi}^2$.  We
include the theoretical uncertainties quoted in Ref.~\cite{Benson:2004sg}
in the overall covariance matrix used in the fit.  To minimize the
theoretical uncertainty we only use moments with $\ecut \le 2.0\gev$
and obtain $ m_{b} = 4.46^{ + 0.21}_{ -0.23}\gev$ and $\mu_{\pi}^2 =
0.64^{ +0.39}_{ -0.38}\gev^2$ with a correlation of $\rho = -0.94$.

\section{Conclusions}
\label{sec:conclusions}

We have measured the \bxsg branching fraction and moments of the
photon energy spectrum above several minimum photon
energies.  We find $\mathcal{B} (\bxsg)= (3.66 \pm 0.85_{\rm stat} \pm 0.60_{\rm syst})
\times 10^{-4}$ for photon energies \eg above 1.9\gev.
Dividing by an extrapolation 
factor of $0.936 \pm 0.010$~\cite{Buchmuller:2005zv} we obtain
$\mathcal{B} (\bxsg) = (3.91 \pm 0.91_{\rm stat} \pm 0.64_{\rm syst})
\times 10^{-4}$ for $E_{\gamma} > 1.6\gev$.
The moments of the spectrum can be used to improve
the knowledge of the heavy quark parameters $m_b$ and $\mupisq$;
we obtain $ m_{b} = 4.46^{ + 0.21}_{ -0.23}\gev$ and 
$\mu_{\pi}^2 = 0.64^{ +0.39}_{ -0.38}\gev^2$ in the kinetic scheme.  
In addition we measured the isospin asymmetry 
$\Delta_{0-} = -0.06 \pm  0.15_{\rm stat} \pm  0.07_{\rm syst}$
and direct \CP asymmetry 
$\acp = 0.10 \pm  0.18_{\rm stat} \pm  0.05_{\rm syst}$
for photon energies above 2.2\gev.
The full reconstruction (recoil) method provides an almost background free
measurement above photon energies of 2.2\gev. 
Although statistics are limited at present,
this approach is expected to provide a competitive
measurement of the decay
\bxsg with the larger data sample that is being accumulated
at the \B-Factories, in particular as the main systematic uncertainties
will also be reduced with a larger data sample.

% Input the acknowledgements file
We are grateful for the excellent luminosity and machine conditions
provided by our \pep2\ colleagues, 
and for the substantial dedicated effort from
the computing organizations that support \babar.
The collaborating institutions wish to thank 
SLAC for its support and kind hospitality. 
This work is supported by
DOE
and NSF (USA),
NSERC (Canada),
CEA and
CNRS-IN2P3
(France),
BMBF and DFG
(Germany),
INFN (Italy),
FOM (The Netherlands),
NFR (Norway),
MES (Russia),
MEC (Spain), and
STFC (United Kingdom). 
Individuals have received support from the
Marie Curie EIF (European Union) and
the A.~P.~Sloan Foundation.

\begin{table*}[p]
\begin{center}
\caption[]{\label{tab:bfresults} Results for $\mathcal{B}(\bxsg)$ for different minimum photon energies 
with statistical and 
systematic errors. Details on the major contributions to the systematic 
uncertainties are also given.\\}
\begin{tabular}{c c c c c c c c }
\hline\hline
\multicolumn{8}{c}{ $\mathcal{B}(\bxsg)$  $(10^{-4})$}\\
 $E_{\gamma}$ range & Value  &  $\sigma_{\rm stat}$ & $\sigma_{\rm  syst}$ & Background & $\mes$ fit  & Detector    & \bxsg \\
  (\gev)           &        &                      &                      & modeling  & parameterization  & response    &  model \\ \hline
1.9-2.6       &       3.66   &       0.85    &       0.60    &       0.35    &       0.45    &       0.18    &       0.08     \\
2.0-2.6       &       3.39   &       0.64    &       0.47    &       0.31    &       0.34    &       0.07     &      0.06     \\
2.1-2.6       &       2.78   &       0.48    &       0.35    &       0.22    &       0.24    &       0.08     &      0.05     \\
2.2-2.6       &       2.48   &       0.38    &       0.27    &       0.14    &       0.19    &       0.10     &      0.05     \\
2.3-2.6       &       2.07   &       0.30    &       0.20    &       0.10    &       0.15    &       0.04     &      0.05     \\\hline\hline
\end{tabular}
\end{center}
\end{table*}

\begin{table*}[p]
\begin{center}
\caption[]{\label{correlpbf}{Correlations between the systematic uncertainties for the differential decay rate measurements.
(The statistical correlations between $E_{\gamma}$ bins are negligible.)\\}}
\begin{tabular}{c|ccccccc}
\hline\hline
$E_{\gamma}$ interval & \multicolumn{7}{c}{$(1/\Gamma_B) (d\Gamma/dE_{\gamma})$}\\ 
 (\gev)          & 1.9-2.0 &  2.0-2.1 &  2.1-2.2 &  2.2-2.3 &  2.3-2.4 &  2.4-2.5 &  2.5-2.6 \\ \hline
      1.9-2.0       &  1.000 & -0.004 &  0.311 &  0.949 & -0.108 &  0.083 &  0.849\\
      2.0-2.1       &        &  1.000 &  0.912 &  0.096 &  0.805 &  0.721 & -0.087\\
      2.1-2.2       &        &        &  1.000 &  0.352 &  0.712 &  0.699 &  0.152\\
      2.2-2.3       &        &        &        &  1.000 &  0.111 &  0.310 &  0.940\\
      2.3-2.4       &        &        &        &        &  1.000 &  0.969 &  0.115\\
      2.4-2.5       &        &        &        &        &        &  1.000 &  0.341\\
      2.5-2.6       &        &        &        &        &        &        &  1.000\\   \hline\hline
\end{tabular}
\end{center}
\end{table*}

\begin{table*}[p]
\begin{center}
\caption[]{\label{correlstat}{Statistical correlations between \meg\ and \vegs\ 
 measurements with different minimum photon energies $E_{\gamma}$.\\}}
\begin{tabular}{cc|ccccc|ccccc}
\hline\hline
\multicolumn{2}{c|}{$E_{\gamma}$ range} & \multicolumn{5}{c|}{\meg}& \multicolumn{5}{c}{\vegs}\\ 
\multicolumn{2}{c|}{($\gev$)} & 1.9-2.6 &  2.0-2.6 &  2.1-2.6 &  2.2-2.6 &  2.3-2.6 &  1.9-2.6 &  2.0-2.6 &
    2.1-2.6 &  2.2-2.6 &  2.3-2.6  \\ \hline
      &1.9-2.6& 1.000 & 0.503 & 0.195 & 0.053 & -0.001 &  -0.897 & -0.419 & -0.180 & -0.040 & 0.041\\
      &2.0-2.6&       & 1.000 & 0.418 & 0.141 &  0.027 &  -0.310 & -0.807 & -0.368 & -0.093 & 0.069\\
\meg  &2.1-2.6&       &       & 1.000 & 0.427 &  0.157 &  -0.124 & -0.342 & -0.822 & -0.244 & 0.109\\
      &2.2-2.6&       &       &       & 1.000 &  0.408 &  -0.017 & -0.054 & -0.153 & -0.550 & 0.200\\
      &2.3-2.6&       &       &       &       &  1.000 &   0.008 &  0.019 &  0.032 &  0.095 & 0.363\\\hline
      &1.9-2.6&       &       &       &       &        &   1.000 &  0.266 & -0.041 & -0.048 & 0.054\\
      &2.0-2.6&       &       &       &       &        &         &  1.000 &  0.094 & -0.052 & 0.113\\
\vegs &2.1-2.6&       &       &       &       &        &         &        &  1.000 &  0.177 & 0.144\\   
      &2.2-2.6&       &       &       &       &        &         &        &        &  1.000 & 0.350\\
      &2.3-2.6&       &       &       &       &        &         &        &        &        & 1.000\\\hline\hline
\end{tabular}
\end{center}
\end{table*}

\begin{table*}[p]
\begin{center}
\caption[]{\label{correlsys}{Systematic correlations between \meg\ and \vegs\ 
 measurements with different minimum photon energies $E_{\gamma}$.\\}}
\begin{tabular}{cc|ccccc|ccccc}
\hline\hline
\multicolumn{2}{c|}{ $E_{\gamma}$ range} & \multicolumn{5}{c|}{\meg}& \multicolumn{5}{c}{\vegs}\\ 
\multicolumn{2}{c|}{($\gev$)} & 1.9-2.6 &  2.0-2.6 &  2.1-2.6 &  2.2-2.6 &  2.3-2.6 &  1.9-2.6 &  2.0-2.6 &
  2.1-2.6 &  2.2-2.6 &  2.3-2.6  \\ \hline
      &1.9-2.6&  1.000  & 0.556 & 0.811 & 0.707 & 0.128 & 0.298 &  0.875 &  0.599 & -0.030 & -0.013\\
      &2.0-2.6&         & 1.000 & 0.946 & 0.667 & 0.751 & 0.903 &  0.511 &  0.943 &  0.775 &  0.724\\
\meg  &2.1-2.6&         &       & 1.000 & 0.765 & 0.653 & 0.741 &  0.771 &  0.932 &  0.553 &  0.546 \\
      &2.2-2.6&         &       &       & 1.000 & 0.527 & 0.412 &  0.777 &  0.731 &  0.307 &  0.416 \\
      &2.3-2.6&         &       &       &       & 1.000 & 0.745 &  0.102 &  0.658 &  0.911 &  0.964\\ \hline
      &1.9-2.6&         &       &       &       &       & 1.000 &  0.253 &  0.841 &  0.868 &  0.770 \\
      &2.0-2.6&         &       &       &       &       &       &  1.000 &  0.656 & -0.059 & -0.019\\
\vegs &2.1-2.6&         &       &       &       &       &       &        &  1.000 &  0.627 &  0.597\\
      &2.2-2.6&         &       &       &       &       &       &        &        &  1.000 &  0.967\\
      &2.3-2.6&         &       &       &       &       &       &        &        &        &  1.000\\\hline\hline
\end{tabular}
\end{center}
\end{table*}

\end{document}